\documentclass[12pt,psfig]{article}
\usepackage{amssymb}
\usepackage{amsmath,amsthm}
\usepackage{amsfonts}
\usepackage{graphicx}
\usepackage{braket}
\usepackage{graphics}
\usepackage{indentfirst}
\numberwithin{equation}{section}

\usepackage{hyperref}

\begin{document}
\begin{center}\Large\textbf{Radiation of a Massless Closed 
String from Interacting D$p$-branes with Background Fields
in the Superstring Theory}
\end{center}
\vspace{0.75cm}
\begin{center}
\large{Hamidreza Daniali and Davoud Kamani}
\end{center}
\begin{center}
\textsl{\small{Department of Physics, Amirkabir University of
Technology (Tehran Polytechnic), Iran \\
P.O.Box: 15875-4413 \\
e-mails: hrdl@aut.ac.ir , kamani@aut.ac.ir \\}}
\end{center}
\vspace{0.5cm}

\begin{abstract}

In the context of the superstring theory,
we calculate the radiation amplitude of a general massless
closed string from the interaction
of two parallel D$p$-branes. The background field
is the constant antisymmetric tensor $B_{\mu\nu}$, and the
branes have been equipped with the different
$U(1)$ gauge fields. For the large distance of the branes,
the Kalb-Ramond emission will be investigated.

\end{abstract}



\newpage

\section{Introduction}

The study of the D-branes with nonzero
background and internal gauge fields, specially
their interactions, reveals a number
of interesting properties in the string theory
\cite{1}-\cite{13}. In the closed string channel,
the interaction between the branes is realized as the
emission/absorption of a closed string between
the branes. A practical and adequate technique for
calculating such interactions
is the boundary state formalism
\cite{2}, \cite{4}, \cite{8}, \cite{14}-\cite{20}.
This method allows us to explore some of the
branes properties
which are not accessible via the other formalisms.

Besides the branes interactions, the closed string
radiation from the branes has been also studied.
One of the most remarkable setups is the
closed string production
from a single unstable brane \cite{21}-\cite{24}.
This sort of radiation has been
investigated in various configurations \cite{25}-\cite{29}.
The closed string radiation from the
interacting branes is another appealing
configuration \cite{30}-\cite{33}.
In this formalism, an appropriate vertex
operator is sandwiched between the boundary states,
which correspond to the branes.

In this paper, we shall concentrate on the
closed string radiation from the
interacting parallel D$p$-branes.
Our branes have been equipped with
the Kalb-Ramond field and the
internal $U(1)$ gauge potentials.
For calculating the amplitude of the radiation, we shall
use the boundary state formalism.
In the large distance of the branes
and in the low energy limit,
the massless closed strings can be usually emitted in the
three distinct physical processes:
the Bremsstrahlung-like process,
which denotes the emission from
each of the interacting branes,
and an emission from the middle region between the
branes \cite{31, 33}.
These results drastically encouraged and motivated us to
investigate the supersymmetric
version of the closed string radiation
from the interacting D$p$-branes.

This paper is organized as follows. In Sec. \ref{200},
the boundary state, corresponding to a D$p$-brane
with the internal and background fields, will be reviewed.
In Sec. \ref{300}, we shall calculate
the radiation amplitude of a general massless
closed string from the interaction of two parallel
dressed D$p$-branes. In Sec. \ref{400},
the amplitude for the Kalb-Ramond
state radiation from the branes with large distance 
will be presented. Our conclusions
have been given in Sec. \ref{500}.

\section{The boundary state}
\label{200}

For constructing the boundary
state, which corresponds to a D$p$-brane with
the background field and internal gauge potential,
we start with the well-known closed string action
\begin{equation}
\label{2.1}
S_\sigma = -\dfrac{1}{4 \pi \alpha^\prime} \int_\Sigma
d^2 \sigma \left(\sqrt{-h} h^{ab} G_{\mu\nu}
+ \epsilon^{ab} B_{\mu\nu} \right)
\partial_a X^\mu \partial_b X^\nu +
\dfrac{1}{2\pi \alpha^\prime} \int_{\partial\Sigma}
d\sigma A_\alpha \partial_\sigma X^\alpha ,
\end{equation}
where $\mu , \nu$ are the spacetime indices,
i.e., $\mu,\nu \in \{0, 1,\cdots, 9\}$,  and
$\alpha, \beta \in \{0, 1,\cdots, p\}$
represent the worldvolume directions of the 
D$p$-brane. For the perpendicular directions to the
worldvolume, we shall apply $i,j \in \{p+1, \cdots, 9\}$.
We use the flat spacetime with the
metric $G_{\mu\nu}= {\rm diag}(-1, 1,\cdots, 1)$,
flat worldsheet
$h_{ab} =\eta_{ab}$ with $a, b \in \{\tau, \sigma\}$,
and a constant Kalb-Ramond field $B_{\mu\nu}$.
Besides, we adopt the Landau gauge $A_\alpha = -\frac{1}{2}
F_{\alpha\beta} X^\beta$ with the constant
field strength $F_{\alpha\beta}$.

The equation of motion and the boundary state equations
are derived by vanishing the variation of the action.
Consequently, using the coherent state approach \cite{34} and
the closed string mode expansion,
the oscillating part of the bosonic
portion of the boundary state takes the form
\begin{eqnarray}
\label{2.2}
|B_x\rangle_{\rm osc} &=&
\sqrt{- \det (\mathbf{1}-\mathcal{F})}
\exp \left[ - \sum_{m=1}^{\infty} \left(\dfrac{1}{m}
\alpha^\mu_{-m} {S}_{\mu \nu}
{\tilde \alpha}^{\nu}_{-m}\right)\right]|0 \rangle,
\label{2.11}
\end{eqnarray}
in which, we defined
\begin{equation}\label{2.3}
{S}^{\mu}_{\ \nu} = \left\{Q^{\alpha}_{\ \beta}
\equiv \left(\dfrac{\mathbf{1}+\mathcal{F}}
{\mathbf{1} - \mathcal{F}}\right)^\alpha_{\;\;\;\beta}
,-\delta^{i}_{\;j}\right\}, \qquad
\mathcal{F}_{\alpha\beta} \equiv F_{\alpha\beta}
- B_{\alpha\beta}.
\end{equation}
The normalization factor
$\sqrt{- \det (\mathbf{1}-\mathcal{F})}$
comes from the disk partition function \cite{7}.
The zero-mode part of the boundary state
possesses the feature
\begin{equation}
\label{2.4}
|B_x\rangle^{(0)}= \int_{-\infty}^{+\infty}
\prod_{i = p+1}^{9} \left[\dfrac{d P^i}{2\pi}
\exp\left(-i P^i y^i\right) |P^i\rangle\right]
\prod_\alpha |P^\alpha=0\rangle.
\end{equation}
The momentum $P^\mu$ has nonzero components only
in the transverse directions.

For receiving the boundary state equations of the
worldsheet fermions, we should apply the supersymmetry
transformations on the bosonic boundary state equations.
Accordingly, the following substitutions should be exerted
in the bosonic boundary state equations
\begin{eqnarray}
\partial_+ X^\mu (\sigma, \tau)
\rightarrow  - i \eta \psi^\mu_+(\sigma, \tau),
\qquad
\partial_- X^\mu (\sigma, \tau)
\rightarrow   \psi^\mu_-(\sigma, \tau),\label{2.5}
\end{eqnarray}
where $ \partial_\pm \equiv \frac{1}{2}
(\partial_\tau \pm \partial_\sigma)$,
and $\eta= \pm 1$ will be used for applying the
GSO projection on the boundary states.

Let us denote the fermionic oscillators
in the R-R (NS-NS) sector with $d_m^\mu$ ($b_r^\mu$).
Thus, we acquire
\begin{equation}
|B_\psi,\eta\rangle_{\rm NS}
= \exp \left(i\eta \sum_{r=1/2}^\infty
b^\mu_{-r} S_{\mu\nu} \tilde{b}_{-r}^\nu\right)
|0\rangle_{\rm NS},\label{2.6}
\end{equation}
which represents the fermionic part of the NS-NS sector
boundary state, and
\begin{eqnarray}
|B_\psi,\eta\rangle_{\rm R}=
\dfrac{1}{\sqrt{-\det (\mathbf{1}- \mathcal{F}})}
\exp \left(i\eta \sum_{m=1}^\infty d^\mu_{-m} S_{\mu\nu}
\tilde{d}^\nu_{-m}\right) |B_\psi,\eta\rangle_{\rm R}^{(0)},
\label{2.7}
\end{eqnarray}
is the fermionic part of the R-R sector boundary state.
For both type IIA and type IIB theories,
the explicit form of the zero-mode portion of the boundary
state is given by
\begin{equation}
|B_\psi,\eta\rangle_{\rm R}^{(0)} =
\left(C \Gamma^0 \cdots \Gamma^p \;
\dfrac{1+i\eta \Gamma^{11}}
{1+i\eta} \mathcal{G}\right)_{AB} |A\rangle
\otimes |\tilde{B}\rangle, \label{2.8}
\end{equation}
where $A$ and $B$ denote the 32-dimensional indices for the
$\Gamma$-matrices and spinors in the 10-dimensional spacetime,
$|A\rangle \otimes |\tilde{B}\rangle$
is the vacuum of the zero modes
$ d_0^\mu$ and $\tilde{d}_0^\mu $, and $C$
is the charge conjugate matrix.
Furthermore, the matrix $\mathcal{G}$ must satisfy
the equation
$\Gamma^\alpha\mathcal{G}=
Q^\alpha_{\ \beta}\; \mathcal{G} \Gamma^\beta$.
Hence, this matrix possesses the following
conventional solution
\begin{equation}
\mathcal{G} =  *\exp \left(\frac{1}{2}
\mathcal{F}_{\alpha\beta} \Gamma^\alpha \Gamma^\beta \right)* .
\label{2.9}
\end{equation}
The notation $* \ *$ indicates that the exponential should
be expanded such that all $ \Gamma$-matrices anticommute.
Thus, we have a finite number of terms.

The following direct product exhibits the total boundary state,
associated with the D$p$-brane, for both the NS-NS
and R-R sectors
\begin{equation}
|B , \eta\rangle_{\rm NS(R)}
= \frac{T_p}{2} |B_x\rangle \otimes
|B^{\rm gh}\rangle \otimes |B_\psi
,\eta\rangle_{\rm NS(R)} \otimes
|B^{\rm sgh} ,\eta\rangle_{\rm NS(R)},
\label{2.10}
\end{equation}
where $T_p$ is the brane tension.
Note that the background
and internal fields have no effect on the ghost and
superghost parts of the boundary states.
Since the explicit expressions of
$|B^{\rm gh}\rangle$ and
$|B^{\rm sgh} ,\eta\rangle_{\rm NS(R)}$
are available in the literature,
we shall not write them here.

\section{The closed string radiation}
\label{300}

In the closed string channel, two D-branes interact by
exchanging a closed string between two boundary states,
associated with the D-branes.
Let $\sigma$ be the periodic coordinate of the
string worldsheet, i.e. $0 \le \sigma\le\pi$,
and $0 \le \tau \le t$ be its length coordinate.
Besides, the fields on the two D$p$-branes are labeled
with the subscripts 1 and 2.
In addition, we shall apply the momenta $k_1$ and $k_2$
for the emitted closed string from the
first brane and the absorbed one by
the second brane, respectively.
The radiation of a closed string state
via the interacting branes is elaborated by inserting
the associated vertex operator of the string state,
$V(\sigma, \tau)$, into the interaction amplitude
\begin{equation}
\label{3.1}
\mathcal{A}_{\rm NS(R)}(\eta_1 , \eta_2)
= \int_{0}^{\infty} {\rm d}t \int_{0}^t {\rm d}\tau 
\int_{0}^\pi {\rm d}\sigma \;
_{\rm NS(R)}\langle B^1 , \eta_1|
e^{-tH_{\rm NS(R)}} V(\sigma , \tau)
|B^2, \eta_2\rangle_{\rm NS(R)},
\end{equation}
where $H_{\rm NS}$ ($H_{\rm R}$)
represents the total closed
string Hamiltonian in the NS-NS (R-R) sector.
Note that, because of the effects of the GSO projection,
the total amplitude is given by
the summation over the spin structures.

Let us define $z \equiv \sigma +i \tau$ and
$\partial \equiv \partial_z$.
The vertex operator in Eq. \eqref{3.1} for a general
massless closed string has the following form
\begin{eqnarray}
\label{3.2}
V(z, \bar{z}) = \epsilon_{\mu\nu}\left(
\partial X^\mu(z, \bar z) - \dfrac{1}{2}p\cdot \psi(z)
\psi^\mu(z)\right)\left( \bar{\partial}
X^\nu(z, \bar z) + \dfrac{1}{2} p\cdot \tilde{\psi}(\bar z)
\tilde{\psi}^\nu(\bar z)\right) e^{ip\cdot X(z, \bar z)},
\end{eqnarray}
in which $\epsilon_{\mu\nu}$ denotes the polarization tensor,
and $p$ (with $p^2=0$) indicates the momentum of the radiated
massless closed string.
For simplicity, similar to the Refs. \cite{31},
\cite{35} and \cite{36}, we
assume that the polarization tensor
has non-vanishing elements only
in the perpendicular directions
of the brane worldvolume, i.e., $\epsilon_{\alpha\beta} 
= \epsilon_{\alpha i} = \epsilon_{i\alpha}=0$.

After heavy calculations,
the radiation amplitude finds the following feature
\begin{eqnarray}
\label{3.3}
\mathcal{A}_{\rm NS(R)}(\eta_1 , \eta_2)
&=&\epsilon_{ij} \left[ \prod_{\alpha=0}^{p}
\delta(p^\alpha)\right]
\int_0^\infty {\rm d}t'\int_0^\infty {\rm d}\tau
\int_{0}^\pi {\rm d}\sigma
\int_{-\infty}^{+\infty} \prod_{i= p+1}^9
\frac{{\rm d}k_1^i}{2\pi} e^{(t'+\tau)\xi} e^{i k_1^i b^i}
\nonumber \\
&\times &e^{-t'\alpha'k_1^2} e^{-\alpha' \tau k_2^2} \
\mathcal{Z}^x \ \mathcal{Z}^\psi_{\rm NS(R)}(\eta_1 , \eta_2) \
\langle e^{i p\cdot x}\rangle  \
\mathcal{M}^{ij}_{\rm NS(R)},
\end{eqnarray}
where $\xi = 0 \ (-5/2)$ has been obtained by
regularizing the Hamiltonian in the R-R (NS-NS) sector.
Using the notation
\begin{eqnarray}
\langle \mathcal{O}(\sigma , \tau)
\rangle_{\rm osc}
\equiv \dfrac{_{\rm osc}\langle B^1 , \eta_1|
e^{-t H_{\rm osc} }\mathcal{O} (\sigma , \tau)
|B^2 , \eta_2 \rangle_{\rm osc}}{_{\rm osc}
\langle B^1, \eta_1| e^{-t H_{\rm osc} }
|B^2, \eta_2\rangle_{\rm osc}},
\nonumber
\end{eqnarray}
the matrix elements $\mathcal{M}^{ij}_{\rm NS(R)}$
take the form
\begin{eqnarray}
\label{3.4}
\mathcal{M}^{ij}_{\rm NS(R)} &=&
\langle \partial X^i \bar\partial X^j\rangle_{\rm osc}
- \langle \partial X^i p.X\rangle_{\rm osc} \langle
\bar \partial X^j p\cdot X\rangle_{\rm osc}
-\alpha'^2 k_1^i k_1^j
\nonumber\\
&+& \frac{1}{4} \Big(\langle p\cdot
\psi \ p\cdot \tilde \psi\rangle
\langle\psi^i \tilde\psi^j\rangle
- \langle p\cdot \psi \ \psi^i \rangle
\langle p\cdot \tilde \psi \ \tilde\psi^j\rangle
+ \langle p\cdot \tilde\psi \  \psi^i
\rangle \langle p\cdot \psi \
\tilde\psi^j\rangle\Big)_{\rm NS(R)}
\nonumber \\
&+&\frac{i}{2} \Big( \langle \partial
X^i p\cdot X\rangle_{\rm osc}
\langle p\cdot \tilde \psi \ \tilde\psi^j\rangle_{\rm NS(R)}
- \langle \partial X^j p\cdot X\rangle_{\rm osc}
\langle p\cdot \psi\ \psi^i\rangle_{\rm NS(R)}\Big)
\nonumber\\
&-& \alpha' k^i \Big(i
\langle \bar \partial X^j p\cdot X\rangle_{\rm osc}
+ \frac{1}{2} \langle p\cdot \tilde\psi \
\tilde \psi^j\rangle_{\rm NS(R)}\Big)
\nonumber \\
&+& \alpha' k^j \Big(i
\langle \partial X^i p\cdot X\rangle_{\rm osc}
- \frac{1}{2} \langle p\cdot \psi \
\psi^i\rangle_{\rm NS(R)}\Big).
\end{eqnarray}
The partition functions $\mathcal{Z}^x$
and $\mathcal{Z}^\psi_{\rm NS(R)}(\eta_1 , \eta_2)$
are specified by
\begin{eqnarray}
\mathcal{Z}^x &=& \ _{\rm osc}\langle B^1_x|
e^{-t H_{\rm osc}}|B^2_x\rangle_{\rm osc}\;
\langle B^{\rm gh}|e^{-t H_{\rm gh}}|B^{\rm gh}\rangle,
\label{3.5}\\
\mathcal{Z}^\psi_{\rm NS(R)} (\eta_1 , \eta_2)
&=& \Bigg(\ _{\rm osc}\langle B^1_\psi,\eta_1|
e^{-t H_{\rm osc}}|B^2_\psi ,\eta_2\rangle_{\rm osc}\;
\langle B^{\rm sgh} , \eta_1|e^{-t H_{\rm sgh} }
|B^{\rm sgh}, \eta_2\rangle \Bigg)_{\rm NS(R)}.
\label{3.6}
\end{eqnarray}
In the amplitude \eqref{3.4}, we defined the impact
parameter $b^i = y^i_1 - y^i_2$. Besides,
by using the Wick's rotation
$\tau \to -i\tau$, we introduced another
proper time $t' = t - \tau$.
Note that, the presence of the factor
$\prod_{\alpha=0}^{p} \delta(p^\alpha)$
manifestly elaborates that the
radiated closed string moves perpendicular to the brane.

\subsection{The correlators and partition functions}
\label{301}

Given the explicit form of the Hamiltonian and 
the boundary states \eqref{2.2} and \eqref{2.4}, 
the bosonic correlators and partition functions
take the features 
\begin{equation}
\label{3.7}
\mathcal{Z}^x =
\sqrt{\det\Big[(\mathbf{1}- \mathcal{F}_1) (\mathbf{1}
- \mathcal{F}_2)\Big]} \prod_{n=1}^{\infty}
\bigg{\{}\left[{\rm det} \left(\mathbf{1}
- {\mathcal{S}} \ q^{2n}\right)\right]^{-1}
\left({\mathbf 1}- q^{2n} \right)^{2}\bigg{\}},
\end{equation}
\begin{eqnarray}
\langle \partial X^\mu X^\nu \rangle_{\rm osc}
&=& i \alpha^\prime \Bigg\{\eta^{\mu\nu}
\sum_{n=1}^{\infty} \text{Tr} \left( \dfrac{{\mathbf 1}
+ {\mathcal{S}} q^{2n} }{{\mathbf 1} - {\mathcal{S}}
q^{2n}}\right)  - \sum_{n=1}^{\infty}
\left(S_2 S_1 \right)^{\nu\mu}
\text{Tr} \left( \dfrac{ {\mathcal{S}}
q^{2n} }{{\mathbf 1} -{\mathcal{S}} q^{2n}}\right)
\nonumber\\
&-& \sum_{n=0}^{\infty} S^{ \mu\nu}_{2}
\text{Tr} \Bigg[ \dfrac{{\mathcal{S}} q^{2n} e^{-4t'}}
{{\mathbf 1}- {\mathcal{S}} q^{2n} e^{-4t'}}
\left( {\mathbf 1} + \dfrac{1}{{\mathbf 1}- {\mathcal{S}}
q^{2n} e^{-4t'}}\right)\Bigg]
\nonumber\\
&+& \sum_{n=0}^{\infty}\left(S_{1}\right)^{\nu\mu}
\text{Tr} \Bigg[ \dfrac{{\mathcal{S}} q^{2n} e^{-4\tau}}
{{\mathbf 1}- {\mathcal{S}} q^{2n} e^{-4\tau}}
\left({\mathbf 1} + \dfrac{1}
{{\mathbf 1}- {\mathcal{S}} q^{2n} e^{-4\tau}}
\right)\Bigg]\Bigg\} ,
\label{3.8}
\end{eqnarray}
\begin{eqnarray}
\langle e^{ip\cdot X_{\rm osc}} \rangle
& = & \prod_{n=1}^{\infty}
\det \left({\mathbf 1}- {\mathcal{S}}
q^{2n}\right)^{\frac{\alpha^\prime}{2n}
p_\mu p_\nu \left(S_2S_1\right)^{\mu\nu}}
\nonumber\\
&\times& \prod_{n=0}^{\infty}
\det\left[ \left({\mathbf 1}- {\mathcal{S}} q^{2n} e^{-4t'}
\right)^{p_\mu p_\nu S^{ \mu\nu}_{2}}\left({\mathbf 1}
- {\mathcal{S}} q^{2n} e^{-4\tau}
\right)^{p_\mu p_\nu S_1^{\mu\nu}}
\right]^{-\frac{\alpha^\prime}{2(n+1)}}
\nonumber\\
&\times& \prod_{n=0}^{\infty} \det \left\{ \exp
\left[ \frac{\alpha^\prime p_\mu p_\nu
S^{ \mu\nu}_{2}}{2(n+1)}  \left({\mathbf 1}
- {\mathcal{S}} q^{2n}e^{-4t'}\right)^{-1} \right] \right\}
\ \ \
\nonumber\\
&\times& \prod_{n=0}^{\infty} \det \left\{ \exp \left[
\frac{\alpha^\prime p_\mu p_\nu S_1^{\mu\nu}}{2(n+1)}
\left({\mathbf 1}- {\mathcal{S}} q^{2n}
e^{-4\tau}\right)^{-1} \right]\right\},
\label{3.9}
\end{eqnarray}
where we defined $\mathcal{S}
= S_1^{\rm T} S_2 $ and $ q = e^{-2t}$.
Note that $\langle {\bar \partial} X^\mu  X^\nu
\rangle_{\rm osc}=-\langle {\partial} X^\mu X^\nu
\rangle_{\rm osc}$. Using the equation of
motion $\partial \bar\partial X =0$,
the two-derivative correlator can be obtained from
Eq. \eqref{3.8}, by
taking the derivative with respect to $\bar z$.

Now consider the fermionic section.
For the NS-NS sector, by employing the boundary
state \eqref{2.6}, we receive
\begin{eqnarray}
\label{3.10}
\mathcal{Z}^\psi_{\rm NS}(\eta_1 , \eta_2) =
\prod_{n=1}^\infty \bigg\{\left[\det \left(\mathbf{1}
+ \eta_1\eta_2 \ \mathcal{S} q^{2n-1}\right)\right]
\left(1+ \eta_1 \eta_2 \ q^{2n-1}\right)^{-2}\bigg\},
\end{eqnarray}
\begin{eqnarray}
\label{3.11}
\langle\psi^\mu \psi^\nu\rangle_{\rm NS}
= \langle\tilde\psi^\mu \tilde\psi^\nu\rangle_{\rm NS}
&=& \sum_{n=1}^\infty \left\{ \eta^{\mu\nu} {\rm Tr}
\left(\frac{1}{\mathbf{1} - \mathcal{S} q^{2n}}\right)\right.
\nonumber \\
&-&\left. (-\eta_1\eta_2)^n
(S_2S_1)^{\nu\mu} {\rm Tr}
\left(\frac{\mathcal{S} q^{n}}{\mathbf{1}
- \mathcal{S} q^{2n}}\right)\right\},
\end{eqnarray}
\begin{eqnarray}
\langle \psi^\mu \tilde\psi^\nu\rangle_{\rm NS}
&=& i \eta_1 \sum_{n=0}^\infty (-\eta_1\eta_2)^n
(S_1)^{\nu\mu} {\rm Tr} \left[ \frac{\mathcal{S}
q^n e^{-2\tau}}{\mathbf{1}- \mathcal{S} q^{2n} e^{-4\tau}}
\left(\mathbf{1}
+ \frac{1}{\mathbf{1}
- \mathcal{S} q^{2n} e^{-4\tau}}\right)\right]
\nonumber \\
&+& i \eta_2 \sum_{n=0}^\infty (-\eta_1\eta_2)^n
S_2^{\mu\nu} {\rm Tr} \left[ \frac{\mathcal{S} q^n e^{-2t'}}
{\mathbf{1}-\mathcal{S} q^{2n}
e^{-4t'}} \left(\mathbf{1} + \frac{1}{\mathbf{1}
- \mathcal{S} q^{2n}e^{-4t'}}\right)\right].
\label{3.12}
\end{eqnarray}
The correlator 
$\langle \tilde\psi^\mu \psi^\nu\rangle_{\rm NS}$
can be obtained from
$\langle \psi^\mu \tilde\psi^\nu\rangle_{\rm NS}$ via the
replacements $S_1 \to S^{\rm T}_1$ and $S_2 \to S^{\rm T}_2$.

Because of the zero-mode part of the R-R sector,
the computations in this sector require specific treatment.
Using Eqs. \eqref{2.7} and \eqref{2.8},
the partition function finds the form 
\begin{eqnarray}
\mathcal{Z}_R^\psi (\eta_1 , \eta_2)
&=& \frac{(-1)^p}{\sqrt{\det\Big[(\mathbf{1}
-\mathcal{F}_1) (\mathbf{1}- \mathcal{F}_2)\Big]}}
\prod_{n=1}^\infty \bigg\{\left[\det  \left(\mathbf{1}
+ \eta_1\eta_2 \ \mathcal{S} q^{2n}\right)\right]
\left(1+ \eta_1 \eta_2 \ q^{2n}\right)^{-2}\bigg\}
\nonumber \\
&\times & \left( i  \mathbf{T}'\delta_{\eta_1\eta_2 ,1}
- \frac{1}{2} \mathbf{T} \delta_{\eta_1\eta_2 ,-1}\right),
\label{3.13}
\end{eqnarray}
in which $\mathbf{T}$ and $\mathbf{T'}$ are defined by
\begin{eqnarray}
\mathbf{T}  \equiv  {\rm Tr} \left(\mathcal{G}_1 C^{-1}
\mathcal{G}_2^T C\right),
\qquad
\mathbf{T}' \equiv {\rm Tr} \left(\mathcal{G}_1 C^{-1}
\mathcal{G}_2^T C \Gamma_{11}\right).
\label{3.14}
\end{eqnarray}

The correlators in the R-R sector are separated
into the oscillatory part and zero-mode portion, i.e,
$\langle \mathcal{O} \rangle_{\rm R}
=\langle \mathcal{O} \rangle_{\rm R}^{\rm osc}
+ \langle \mathcal{O} \rangle_{\rm R}^{(0)} $.
For the oscillatory part, one finds
\begin{eqnarray}
\label{3.15}
\langle\psi^\mu \psi^\nu\rangle_{\rm R}^{\rm osc}
= \langle\tilde\psi^\mu \tilde\psi^\nu\rangle_{\rm R}^{\rm osc}
&=& \sum_{n=1}^\infty \left\{ \eta^{\mu\nu} {\rm Tr}
\left(\frac{1}{\mathbf{1} - \mathcal{S} q^{2n}}\right)\right.
\nonumber \\
&-&\left. (-\eta_1\eta_2)^n (S_2S_1)^{\nu\mu} {\rm Tr}
\left(\frac{\mathcal{S} q^{2n}}{\mathbf{1}
- \mathcal{S} q^{2n}}\right)\right\},
\end{eqnarray}
\begin{eqnarray}
\langle \psi^\mu \tilde\psi^\nu\rangle_{\rm R}^{\rm osc}
&=& i \eta_1 \sum_{n=0}^\infty (-\eta_1\eta_2)^n
(S_1)^{\nu\mu} {\rm Tr} \left[ \frac{\mathcal{S}
q^{2n} e^{-4\tau}}{\mathbf{1}- \mathcal{S} q^{2n} e^{-4\tau}}
\left(\mathbf{1} + \frac{1}{\mathbf{1} - \mathcal{S}
q^{2n} e^{-4\tau}}\right)\right]
\nonumber \\
&+& i \eta_2 \sum_{n=0}^\infty (-\eta_1\eta_2)^n
S_2^{\mu\nu} {\rm Tr} \left[ \frac{\mathcal{S} q^{2n}
e^{-4t'}}{\mathbf{1}-
\mathcal{S} q^{2n} e^{-4t'}} \left(\mathbf{1}
+ \frac{1}{\mathbf{1} - \mathcal{S}
q^{2n} e^{-4t'}}\right)\right].
\label{3.16}
\end{eqnarray}
The correlator
$\langle \tilde\psi^\mu \psi^\nu\rangle_{\rm R}$
can be obviously obtained from
$\langle \psi^\mu \tilde\psi^\nu\rangle_{\rm R}$ via the
replacements $S_1 \to S^{\rm T}_1$ and $S_2 \to S^{\rm T}_2$.
For the zero-mode part, we obtain
\begin{eqnarray}
\langle d_0^\mu d_0^\nu \rangle_{\rm R}^{(0)}
&=& (-1)^{p+1} \langle \tilde d_0^\mu \tilde d_0^\nu
\rangle_{\rm R}^{(0)}= \frac{(-1)^p}{2}
\left( i \delta_{\eta_1\eta_2,1}
\frac{\mathcal{T}^{\prime \ \mu\nu}}{\mathbf{T}'}
- \delta_{\eta_1\eta_2, -1}
\frac{\mathcal{T}^{\mu\nu}}{\mathbf{T}}\right),
\label{3.17}\\
\langle \tilde d_0^\mu d_0^\nu \rangle_{\rm R}^{(0)}
&=& - \langle d_0^\mu \tilde d_0^\nu \rangle_{\rm R}^{(0)}
= \frac{(-1)^p}{2} \left( i \delta_{\eta_1\eta_2,1}
\frac{\mathcal{T}^{\mu\nu}}{\mathbf{T}'}
- \delta_{\eta_1\eta_2, -1}
\frac{\mathcal{T}^{\prime \ \mu\nu}}
{\mathbf{T}}\right),
\label{3.18}
\end{eqnarray}
where the new tensors are defined as follows 
\begin{eqnarray}
\label{3.19}
\mathcal{T}^{\mu\nu} & \equiv &
{\rm Tr} \left(\mathcal{G}_1 \Gamma^\nu C^{-1}
\mathcal{G}_2^T C \Gamma^p \cdots \Gamma^1
\Gamma^\mu \Gamma^1 \cdots \Gamma^p \right),
\nonumber \\
\mathcal{T}^{\prime \ \mu\nu} & \equiv &
{\rm Tr} \left(\mathcal{G}_1 \Gamma^\nu C^{-1}
\mathcal{G}_2^T C \Gamma^p \cdots \Gamma^1
\Gamma^\mu \Gamma^1 \cdots \Gamma^p \Gamma_{11}\right).
\end{eqnarray}

According to Eqs. \eqref{3.7}-\eqref{3.19},
the integrand of Eq. \eqref{3.3}
merely depends on $\tau$ and $t'$,
and it is independent of the variable $\sigma$. 
Hence, the integration over the worldsheet coordinate
$\sigma$ is given by an overall factor $\pi$.

By introducing all correlators and partition functions
into Eq. \eqref{3.3}, we receive the finalized form of
the partial amplitudes
$\mathcal{A}_{\rm NS(R)}(\eta_1 , \eta_2)$.
The total amplitude is given by
\begin{eqnarray}
&~&\mathcal{A}_{\rm total}= \mathcal{A}(+,+)+
\mathcal{A}(-,-) + \mathcal{A}(+,-)
+ \mathcal{A}(-,+),
\nonumber \\
&~&\mathcal{A}(\eta_1 , \eta_2)
=\mathcal{A}_{\rm NS}(\eta_1 , \eta_2)
+\mathcal{A}_{\rm R}(\eta_1 , \eta_2).
\end{eqnarray}
Due to the very long length of the total amplitude,
we do not explicitly write it.

\section{The emission of the Kalb-Ramond state from
large distance D$p$-branes}
\label{400}
 
In the case of large distance branes, which
corresponds to a long enough time, the exchange of the
massless states extremely possesses the dominant
contribution to the interaction. 
However, in the interaction amplitude the limit 
$t \to \infty$ should merely exert 
on the contribution of the oscillators, 
i.e, on the correlators and partition 
functions. Besides, we should mention that the 
exponent of the modular parameter $q$ in the R-R sector 
is not in the same order as in the NS-NS sector 
(see Eqs. \eqref{3.12} and \eqref{3.16}).
Therefore, in both sectors we shall apply  
the order $\mathcal{O}(q^{2n})$. 

In fact, even in the eikonal approximation \cite{31,32},
in which the recoils of the branes are ignored,
the particle creation quantum-mechanically is permitted.
This fact becomes more obvious when one considers the 
emission of the massless strings in the limit of small 
energy and large inter-brane separation. 
Thus, we shall concentrate on the
lowest order of the energy of the radiated string
and the large distance of the interacting branes.

To compare our results with   
Ref. \cite{31}, we apply the condition
$\mathcal{S} =\mathbf{1}$, which prominently imposes
some restrictions on the parameters of the setup. 
This constraint enables us to recast 
the two-derivative term as a 
combination of the one-derivative terms.
In the calculations we receive a non-integral surface term 
at $\tau , t' = 0$.
Since it has no physical interpretation, it
can be eliminated by assuming $p_\mu p_\nu S_1^{\mu\nu}$,
$p_\mu p_\nu S_2^{\mu\nu}<0$, \cite{31, 32}.
Thus, we obtain
\begin{equation}
\label{4.1}
\langle \partial X^\mu \bar{\partial}
X^\nu \rangle_{\rm osc}|_{t\rightarrow \infty}
=-\langle \partial X^\mu X^\nu
\rangle_{\rm osc}|_{t\rightarrow \infty}
\left[ \langle p\cdot \partial X p\cdot X
\rangle_{\rm osc}|_{t\rightarrow \infty} +
\dfrac{i \alpha^\prime}{2} (k_1^2 - k_2^2)\right].
\end{equation}
In addition, the integration by part gives
the following equivalent relations
\begin{eqnarray}
\label{4.2}
\ \frac{e^{-4(\tau,t')}}{1- e^{-4(\tau,t')}} 
\left( 1 + \frac{1}{1
- e^{-4(\tau,t')}}\right) &\doteq&
\frac{\xi-\alpha'k_{(2,1)}^2}{20\alpha' 
p_\mu p_\nu S_{(1,2)}^{\mu \nu}}.
\end{eqnarray}
These relations reduce the dependency of the trace-terms
to the proper times $t'$ and $\tau$.
Finally, the integrals over the proper
times, in the eikonal approximation, 
is given by \cite{33},
\begin{eqnarray}
\int_0^{\infty} {\rm d}t' \int_0^\infty
{\rm d}\tau \ e^{- \alpha'k_1^2 t'}
e^{- \alpha'k_2^2 \tau}e^{\xi(t'+\tau)}
\langle e^{ip\cdot X_{\rm osc}}
\rangle|_{t\rightarrow\infty}^{{\mathcal{S}}
= \mathbf{1}} \approx \Big[(\alpha' k_1^2 -\xi)
(\alpha' k_2^2 -\xi)\Big]^{-1}.
\label{4.3}
\end{eqnarray}

The polarization tensor of the Kalb-Ramond state has
the properties
$\epsilon^{\rm KB}_{\mu\nu}=-\epsilon^{\rm KB}_{\nu\mu}$ and
$p^\mu \epsilon^{\rm KB}_{\mu\nu}=0$.
Adding all these together yields the amplitude
of the Kalb-Ramond radiation
\begin{eqnarray}
\mathcal{A}^{\rm KB}_{\rm R}|^{\mathcal{S}
=\mathbf{1}}_{t\rightarrow\infty}
&= & \frac{ (-1)^p T_p^2}{128\alpha'^2
(2\pi)^{7-p}} \epsilon_{ij}^{\rm KB}
\left(\prod_{\alpha=0}^p \delta(p^\alpha)\right)
\int_{-\infty}^{+\infty} \prod_{i=p+1}^9
\frac{dk_1^i}{k_1^2 k_2^2} e^{ik_1^ib^i}
\nonumber\\
&\times&\Bigg\{
i \mathbf{T}'^{-1}p_k p_l\left[\mathcal{T}^{ki}
\mathcal{T}^{lj}+ (-1)^{p+1} \mathcal{T}^{ki}
\mathcal{T}^{\prime \ lj}
-\mathcal{T}^{kl} \mathcal{T}^{ij}\right]
\nonumber\\
&-& (2 \mathbf{T})^{-1}p_k p_l\left[ \mathcal{T}^{\prime \ kl}
\mathcal{T}^{\prime \ ij}+(-1)^p\mathcal{T}^{ki}
\mathcal{T}^{ lj}- \mathcal{T}^{\prime \ ki}
\mathcal{T}^{\prime \ lj}\right]
\nonumber\\
&+& 4i\alpha'p_k \Bigg[k_1^i
\left(i\mathcal{T}^{\prime \ kj}
-\frac{1}{2} \mathcal{T}^{kj}\right)
-(-1)^p k_1^j \left(i\mathcal{T}^{\prime \ ki}
-\frac{1}{2} \mathcal{T}^{ki}\right) \Bigg]\Bigg\},
\label{4.4}
\end{eqnarray}
\begin{eqnarray}
\mathcal{A}^{\rm KB}_{\rm NS}|^{\mathcal{S}
=\mathbf{1}}_{t\rightarrow\infty} &=& 0 .
\end{eqnarray}
We observe that,
because of the anti-symmetric property of the polarization
tensor of the Kalb-Ramond state, the NS-NS sector
does not possess any contribution to the radiation amplitude.

Our results, which were acquired
in the eikonal approximation, 
are comparable with the results of Ref. \cite{31}. 
In fact, the emission amplitude 
for any massless state in the large 
distance limit generally includes one of 
the following processes. 
The terms with the propagator 
factors $1/k_1^2$ and $1/k_2^2$ 
elaborate that one of the branes 
emits a massless closed string, 
then it is absorbed by the other brane, 
afterward it moves on the absorber brane for a while, 
the absorber brane finally emits the final 
massless string state. 
This kind of string radiation is similar to the 
bremsstrahlung process. In the emission amplitude,
the term with the factor $1/k_1^2k_2^2$ 
represents a double-pole mechanism, 
in which the massless string radiation 
occurs from the exchanged closed string 
between the two D-branes. 
The latter string obviously causes the interaction 
between the two D-branes. 
 
We observed that the massless string radiation 
in the middle region of the distant branes  
exclusively is for the Kalb-Ramond emission. This 
kind of emission has not occurred in Ref. \cite{31}, 
and not even in the bosonic configurations \cite{33}.
However, the absence of the terms with the propagator factors 
$1/k^2_1$ and $1/k^2_2$ in the radiation amplitude \eqref{4.4}
manifestly demonstrates 
that there is no bremsstrahlung-like process. Precisely, we
do not have any massless string  
radiation from the surfaces of the branes.

\section{Conclusions}
\label{500}

In the first part of the paper, 
we employed the boundary states formalism to
compute the radiation amplitude 
of a general massless closed string
from the interaction of two parallel
D$p$-branes in the context of the superstring theory.
We assumed that the branes have been
dressed with the Kalb-Ramond field
and different $U(1)$ gauge potentials.

In the second part of the paper,
we specifically calculated the radiation amplitude
of the Kalb-Ramond state from the
branes with large distance.
For this case, we imposed some constraints
on the setup parameters. We observed that
the Kalb-Ramond radiation
occurs only in a single physical process, i.e.
the radiation from the exchanged closed string
between the interacting branes.
This radiation exclusively happens in the R-R sector,
and the NS-NS sector does not possess any
contribution to it.


\end{document}